\documentstyle[12pt,psfig,aasms4]{article}
\lefthead{Rosenberg A., et~al. }
\righthead{The metallicity of Palomar 1}

\begin{document}

\title{The metallicity of Palomar 1
\footnote {Based on observations made with the Isaac Newton 
Telescope operated on the island of La Palma by the Isaac Newton Group
in the Spanish Observatorio del Roque de los Muchachos and the IAC-80
Telescope operated on the island of Tenerife by the IAC in the Spanish
Observatorio del Teide, both of the Instituto de Astrof\'{\i}sica de
Canarias.}}

\author{A. Rosenberg}
\affil{Telescopio Nazionale Galileo, Osservatorio Astronomico di Padova, Italy}
\author{G. Piotto, I. Saviane}
\affil{Dipartimento di Astronomia, Universit\`a di Padova, Italy}
\author{A. Aparicio}
\affil{Instituto de Astrof\'{\i}sica de Canarias, Spain}
\author{R. Gratton}
\affil{Osservatorio Astronomico di Padova, Italy}

\begin{abstract}
Palomar~1 is a peculiar galactic globular cluster, suspected to be
younger than the bulk of the Galactic halo objects. However, such a
low age can be confirmed only after a reliable determination of the
metallicity. In the present paper, we use the equivalent widths ($W$)
of the Ca~II triplet on medium resolution spectra in order to
determine the metal content of Pal~1. From the comparison of the
luminosity corrected $W$'s in four stars of Palomar~1 with those of a
sample of stars in each of three calibration clusters (M2, M15, and
M71), we derive [Fe/H]=$-0.6\pm0.2$ on the Zinn \& West (1984) scale
or [Fe/H]=$-0.7\pm0.2$ on the Carretta \& Gratton (1997) scale. We
also obtain a radial velocity $V_r=-82.8\pm 3.3$ Km/s for Pal~1.

\end{abstract}

\section{Introduction\label{sec:intro}}

Palomar~1 is a very faint (M$_V=-2.54$) and peculiar star cluster
discovered by Abell (1955) on the Palomar Sky Survey plates. It is
located about 11.2~kpc from the Sun, 17.3~kpc from the Galactic
center, and 3.6~kpc above the Galactic plane (Rosenberg et~al 1997,
R97).

We have presented a photometric study of Pal~1 in a companion paper
(R97), to which the reader is referred for a summary of the previous
investigations on this object. The most important result from R97 is
that the age of Pal~1 is significantly lower than the bulk of the
Galactic globular clusters (GGC), i.e. 8~Gyr on the scale of Bertelli
et al. (1994, B94). This result critically depends on the assumed
metallicity of the object.

The first determination of the metallicity was obtained by Webbink
(1985). From the correlation between the dereddened giant branch base
colors and the high-dispersion spectroscopic metallicities, he found
[Fe/H]=$-1.01$, adopting (B--V)$_{o,g}=1.08$ and E(B--V)=0.15 from
unpublished data by Da~Costa. However, R97 have shown that the
location of the HB of Pal~1 is rather questionable, making the
estimate of the (B-V)$_{0,g}$ very uncertain. Borissova \& Spassova
(1995) give [Fe/H]=$-0.79$ using the $\Psi$$^2$ parameter of Flannery
\& Johnson (1982) (no error is quoted). This result is based on the
global fit of theoretical tracks to the observed c--m diagram; since
there is no other independent estimate of the distance and age of 
Pal~1, this value of [Fe/H] is to be considered only tentative. In view of
the uncertainties in the previous results, we planned a spectroscopic
investigation of a few Pal~1 stars, in order to have a direct estimate
of its metal content.

As a consequence of the large uncertainties associated to the above
determinations of [Fe/H], and in view of the importance of a reliable
metallicity for an age determination, we investigated if there were
any other possibilities to measure the metal content of Palomar\/1.
As the brightest stars of Pal~1 have an apparent luminosity $V \geq
16.4$ (cf. Figure~\ref{sel_cmd}), a direct determination of the
metallicity with high--resolution spectroscopy is feasible only with
8m--class telescopes. As discussed in R97, photometric methods cannot
provide a good [Fe/H] estimate. We remain with medium--resolution
spectroscopy. The technique of Armandroff \& Da~Costa (1986, 1991) is
perfectly suitable to the case of Pal~1. Their method relies on the
determination of the equivalent widths ($W$) of the Ca~II triplet, and
requires $\sim$2 \AA\ resolution spectra and good photometry.

The observations and the employed reduction techniques are presented
in the following Sect.~\ref{sec:obs}. In Sect.~\ref{sec:abun} we
discuss the metal abundance and radial velocity of Pal~1, resulting
from our spectra. A summary is presented in Sect.~\ref{sec:sum_con}.
A brief account of the photometric observations needed for the
calibration of the metallicity is given in the Appendix.

\section{Observations and reductions\label{sec:obs}}

\subsection{Selection of the targets}

The method proposed by Armandroff and Da Costa (1986, 1991) allows to
determine the metallicity of a sample of cluster stars by comparing
their Ca~II triplet equivalent widths with the corresponding $W$'s of a
set of stars in clusters of known metal content. Due to the high
uncertainty of [Fe/H] for Pal~1, the reference (calibrating) clusters
must cover a large metallicity interval. We chose M2, M15, and M71 as
reference globular clusters (GC). Their metal content is known with
high accuracy: Zinn and West (1984) gives [Fe/H]=$-1.62\pm0.07$ for
M2, $-2.15\pm0.08$ for M15, and $-0.58\pm0.08$ for M71. These
metallicities have been furtherly confirmed by Armandroff
(1989). Images in the $V$ and $I$ bands for these clusters were taken
at the IAC--80 Telescope (see the Appendix) in order to obtain their
color magnitude diagrams (CMD), and the positions and magnitudes of
the bright giants to be used as reference stars.

The stars in each cluster were chosen on the basis of their proximity
to the cluster giant branch in the color-magnitude diagram (see for
example Figure~\ref{sel_cmd} and Figure~\ref{m2_cmd}), although the degree
of crowding and the distance from the cluster center were also used as
selection criteria. We selected 4 stars in Pal~1 (Figure~\ref{sel_cmd}),
and in each of the reference clusters. The observed stars are marked
in Figures~\ref{sel_im}, \ref{m2_im}, \ref{m15_im} and \ref{m71_im}.

\subsection{Observations}

Three long exposure intermediate resolution spectra were obtained for
four stars in the direction of Pal~1, M71, and M2, and three stars in
the direction of M15 at the 2.5m Isaac Newton Telescope (Roque de los
Muchachos Observatory) on September, 7 and 8 1996. The intermediate
resolution spectrograph (IDS) was employed with the 235~mm camera and
an 831~line/mm grating centered at $\lambda$8548 \AA\ to observe the
lines of the Ca~II triplet ($\lambda$8498, $\lambda$8542,
$\lambda$8662 \AA). The detector was a thinned $1024\times1024$
pixel$^2$ Tektronix CCD with a pixel size of 24$\mu$m. The resulting
scale was 1.22 \AA/pix and the instrumental resolution was 2.1 \AA\
(FWHM). A CCD window of $384 \times 1024$ pixels was used to give
390~arcsec of spatial coverage and $\sim$1245 \AA\ of wavelength
coverage. Exposures for the Pal~1 stars ranged from $3
\times 1600$~s to $3 \times 3000$~s and each star exposure was
bracketed by exposures of Cu--Ar--Ne lamps for wavelength
calibration. In order to save as much telescope time as possible,
spectra with two target stars within the slit were taken in each
exposure. Details of the observing log are given in
Table~\ref{tab:obs}.

\subsection{Reductions}

The raw data were reduced to wavelength calibrated sky-subtracted
spectra using standard techniques within IRAF ({\it cf.} Massey et al., 1992).

Fig~\ref{spec} shows the wavelength range covered by the Ca~II triplet
spectra for one star per cluster. The spectra of star III for M71, and
star I for each of the other clusters are shown. The spectra have
been smoothed and normalized to the adopted continuum. They have also
been corrected to the rest frame and the three lines of the Ca~II
triplet are marked by the vertical dotted lines. Since all spectra
are on the same scale, a direct visual comparison of the areas covered
by each line can be made. It is evident that the M71 and Pal~1 $W$'s are
comparable , while those of M2 and M15 are clearly smaller. This
suggests that the metallicities of M71 and Pal~1 are similar (even
taking into account the small correction for the absolute magnitude
effect).
 
The $W$'s of the Ca~II triplet lines at $\lambda$8542 and $\lambda$8662
\AA\ were then determined from the spectra via Gaussian fitting as
discussed by Armandroff \& Da Costa (1991). The line $\lambda$8498 was
not used, since its lower strength in most cases contributes more to
the noise than to the signal. 

In addition to the line--strength
measures, a radial velocity was determined from each stellar spectrum
by measuring the central wavelength of each Ca~II line.

The sum of the equivalent widths is denoted by $W_{8542}+W_{8662}$ and
the values of this line--strength index for all observed stars are
listed in Table~\ref{tab:dat} together with their estimated
uncertainties. The errors were estimated from (a) the uncertainties in
the parameters that define the Gaussian fits which are calculated in
the fitting process, and (b) from the comparison of the individual
measures of the three spectra obtained for each star. The above two
error estimates were giving consistent results.

\section{The radial velocity and metal abundance of Palomar 1.\label{sec:abun}}

\subsection{Radial velocities}

Obtaining the radial velocities of the Pal~1 stars allows both an
estimate of the mean radial velocity of the cluster and an assessment
of the membership. The velocities have been obtained by first
determining the geocentric values and then by correcting for the Earth
motion. The final heliocentric values for the stars of Pal~1 and the
three comparison clusters are presented in Table~\ref{tab:dat},
together with the internal errors estimated from the dispersion in
the three measurements of the Ca~II lines. A brief account of the
procedure we followed is given below.

First, the central wavelength of each Ca~II line was computed using
Gaussian fits, after wavelength calibration. We checked that no
systematic errors were present by comparing the calibration obtained
from the lamp spectra with those obtained from the sky lines which
were present in each stellar spectrum.

As a second step, individual relative heliocentric corrections were
applied within IRAF. These corrections were checked repeating the
calculations with a different package (ESO/MIDAS).

The radial velocity errors reported in Table~\ref{tab:dat} have been
estimated by taking the mean of the three measures (i.e. spectra) for
each star, and evaluating the dispersion. The radial velocities allow
the discrimination between cluster members and non-members by
comparison with published values. We used the data collected in Pryor
and Meylan (1993), who give $v_r = -3.11\pm0.90$ km/s,
$v_r=-107.09\pm0.80$ km/s, and $v_r=-23.16\pm0.24$ km/s for M2, M15
and M71, respectively. In order to discuss the differences between
these values and our estimates, we must also take into account the
internal velocity dispersion of the objects. Pryor and Meylan (1993)
give $\sigma_v=7.39\pm0.64$ km/s, $8.95\pm0.59$ km/s, and
$2.16\pm0.17$ km/s for the same clusters. We have classified the stars
whose velocities deviate by more than $3\,\sigma_v$ from the cluster
mean radial velocity as non-members. Therefore star III of M15 and
stars I,II of M71 were discarded and have not been used in any of the
following calculations. Stars III and IV of M2 have a radial velocity
which is only 1.2 and 1.3 sigma lower than the mean value. The
equivalent widths reported in Table~\ref{tab:dat} are compatible with
those of the other members of the cluster, so, though the membership 
cannot be fully established, these two stars are likely members. 

Therefore, our computations were repeated both
including and excluding these two stars. We did not find any
significant differences in the results, due to the similarity of the
$W'$s.

In any case, the following discussion is made without these uncertain
objects, unless explicitly stated. 

An independent estimate of the measurement errors can be made by
comparing the published radial velocities of M2, M15 and M71 with our
velocities. Excluding the uncertain members III and IV of M2, we
find an almost null zero--point offset and a dispersion of $\simeq
3$~km/s, which is slightly larger than our estimated internal errors
but consistent with the velocity dispersion of the stars in these
galactic globular clusters. If stars III and IV of M2 are included in
the calculations, the dispersion becomes $\simeq5$~km/s (and the
offset would be $-2.3$~km/s). Taking the weighted mean of the radial
velocities in Table~\ref{tab:dat}, we estimate for Pal~1 a radial
velocity $v_r$ = $ -82.8 \pm 3.3$ ~km/s.

\subsection{Abundance} \label{abun}

Da Costa \& Armandroff, (1995,DA95) demonstrated that, for stars in the same cluster
(i.e. metallicity), there is a linear relation between their
magnitudes and the $W$ measured for the Ca~II lines $\lambda$8542 and
$\lambda$8662 \AA. The proposed relation has the following form \[
W_{8542} + W_{8662} = a \cdot (V - V_{\rm HB}) + b \] For $V-V_{\rm
HB} < -0.5$ the value of the slope is $a = -0.62$~\AA~mag$^{-1}$
(Armandroff \& Da Costa 1991, Suntzeff et al. 1993, DA95). The same
authors find a well defined relation between the metallicity of a
cluster and the so-called reduced equivalent width $W'$, which is
defined as $W' = < W_{8542} + W_{8662} +0.62 \cdot (V - V_{\rm HB})>$.

The magnitude interval so far used in order to compute the parameter
$a$ comprises all the stars on the RGB which are brighter than the
HB. In the case of Pal~1, this part of the RGB is absent, and we were
forced to find a new relation using fainter stars. Indeed, Figure~5 of
Suntzeff et al. (1993, S93) shows that the slope of the $W_{8542} +
W_{8662}$ vs. $V-V_{\rm HB}$ relation flattens for stars fainter than
the HB.

The data in Tab.~\ref{tab:dat} were used to calculate
the appropriate value of $a$, according to the following
procedure. Lines with fixed values of $a$ were fitted to the data and the 
corresponding values of the intercepts $b$ were found for each cluster. 
The offsets $b$ were added to each subset, and a new dataset, which 
now has a common zero point, was created. A linear fit was repeated with 
the new dataset, and the RMS value of the dispersion of the data around 
the fit was computed. These operations were repeated for different values 
of $a$, and we accepted the value of $a$ that minimizes the RMS. 
We found $a =-0.4 \pm 0.2$. 

In Figure~\ref{EW} the summed equivalent widths, with their
associated errors, are plotted against $V-V_{HB}$ for the cluster
member stars (including stars III and IV of M2).  This figure
illustrates that the slope $a$ is essentially determined by the M2
data. The value of $a$ is not too different from the value valid for the
commonly used brighter stars, and is entirely compatible with the
trend suggested by the fainter stars (NGC~6397) in Figure~5 of S93.

As in the previously mentioned studies, we can now define a reduced
$W'$ as $W' = < W_{8542} + W_{8662} +0.4 \cdot (V - V_{\rm HB})>$, and
use this $W'$ to determine the metallicity of Pal~1.

Fig~\ref{met} represents the abundance calibration for the Ca~II line
strengths. The values of [Fe/H] from Armandroff (1989), on the Zinn \&
West (1984) scale, are plotted vs. the weighted means of the $W'$ of
the 3 calibration clusters. The relation between these two quantities
changes its slope at [Fe/H]$\sim-1.3$ (see e.g. Figure~3 in DA95), so in
our figure we show linear fits to the data obtained using the three
calibration clusters (dashed line) and just the two more metal rich M2
and M71 (solid line).

We still need the $V-V_{HB}$ values for Pal~1 in order to compute the
$<W'>$ and then its metallicity. It is not easy to determine the
$V-V_{HB}$ difference for the stars of Pal~1 because this cluster has
no HB stars, and it is younger than the comparison clusters. On the
other hand, we know that the $W$'s of Pal~1 are similar to those of M71
(even taking into account the small absolute magnitude corrections).
All the other parameters being similar, this means that the location
of the Pal~1 HB should be close to that of M71. Nevertheless, since
the luminosity of the HB depends on both the cluster age and the
metallicity, we evaluated the reasonable limits within which the Pal~1
HB could vary. Taking the extreme values we will have an estimate of
the maximum error on the metallicity determination.

It has been established that Pal~1 has an age of $\simeq 8 \pm 2$~Gyr
on the Bertelli et al. (1994, B94) scale (Rosenberg et al. 1997),
which is $\sim 8$~Gyr lower than the ``standard'' value often adopted
for the globular cluster ages, on the same scale. According to Eq.~12
of B94, this change in age would make the HB $V$ absolute luminosity
$\sim 0.2$~mag brighter. Hence, taking an absolute value for the HB
luminosity of $0.7 \pm 0.2$~mag for M71 (cf. Appendix~\ref{appendix}),
the corresponding luminosity of the Pal~1 HB is $0.5 \pm 0.2$~mag.
Even taking a variation for the Pal~1 metallicity of $\pm 0.3$~dex
(see below), this would imply a change in the HB luminosity of only
$\sim 0.06$~mag (see again B94). The error on $V_{\rm HB}$ is
therefore almost entirely due to the error in the location of the M71
HB. An absolute magnitude of $0.5$~mag corresponds to an apparent
magnitude $V_{HB} = 16.3 \pm 0.35$ at the distance of Pal~1 (R97),
where the error on the distance modulus has been added to the total
error on $V_{\rm HB}$. It is important to note that, even adopting a
typical GC age for Pal~1, the resulting estimate of its metallicity
would vary by $\sim 0.04$~dex only.

With this value, the weighted mean of the Pal~1 $W$ can finally be
computed and entered into the relations of Figure~\ref{met}. The
resulting metallicity is $[Fe/H] = -0.6 \pm 0.2$, for both relations
previously defined. Using the Carretta \& Gratton (1997) metallicity
scale, the result would be $[Fe/H] = -0.7 \pm 0.2$ (see
Table~\ref{tab:fehpal1}).

The error has been estimated taking into account the following
contributions: the error on the slope for finding the single $W'$ and
their weighted means, the error on the metallicity of the calibration
clusters, and finally, the error in the determination of the
$V-V_{HB}$ value.

\section{Summary \label{sec:sum_con}}

Medium resolution spectra were collected and reduced for a sample of
stars in Pal~1, M2, M15 and M71. We measured the $W$'s for the Ca~II
triplet in each spectra. A linear correlation was found between the $W$'s
and their luminosities, in the form $W_{8542} + W_{8662} = a \cdot (V
- V_{\rm HB}) + b$. The luminosity corrected $W$'s were calibrated as a
function of metallicity by using the stars in M2, M15, and M71.
Applying the same relation to the $W$'s of Pal~1 we obtained
[Fe/H]=$-0.6\pm0.2$ on the Zinn \& West (1984) scale or
[Fe/H]=$-0.7\pm0.2$ on the Carretta \& Gratton (1997) scale.

A reliable estimate of the metallicity of Pal~1 is fundamental for an
estimate of its age, particularly important in view of the peculiar
properties of this cluster (R97). We also measured the heliocentric
radial velocities for all the observed stars. A comparison between the
published radial velocities of M2, M15, and M71, with those of our
sample, has been used to identify the cluster members. Our average
velocities are in agreement with the published ones, excluding any
systematic errors in our measurements. For the first time, we can give
the heliocentric radial velocity of Pal~1: $V_r=-82.8\pm 3.3$ Km/s.

\acknowledgments

This project has been partially supported by the Agenzia Spaziale
Italiana. The observation run has been supported by the European
Commission through the Activity ``Acces to Large-Scale Facilities''
within the Programme ``Training and Mobility of Researchers'',
awarded to the Instituto de Astrofisica de Canarias to fund European
Astronomers access to its Roque de Los Muchachos and Teide
Observatories (European Northern Observatory), in the Canary Islands.
We recognize partial support by the Instituto de Astrofisica de
Canarias (grant P3/94) and by a Spanish-Italian integrated action. We
thank Prof. Jack Sulentic for the careful reading of the manuscript.

\appendix

\section{Observations of M2, M15 and M71.}
\label{appendix}

Three long exposures of M2, M15 and M71 in the $V$ and $I$ bands
(1200s and 900s respectively) were collected with the IAC--80
Telescope, on August 11 and 12, 1996, at the Observatory of Teide in
Tenerife, Canary Islands, Spain. These frames cover the NW quadrant
of each cluster, and were obtained with the aim of selecting the
target stars to be observed in the spectroscopic run. The resulting
fields are shown in Figure~\ref{m2_im}, \ref{m15_im} and
\ref{m71_im}, and the target stars have been marked and numbered.
During the two nights the weather conditions were stable, although the
seeing, on average, was poor (1.8\arcsec-2.1\arcsec). 

The camera was equipped with an EEV CCD at the Cassegrain focus, and
the resulting scale was $0\farcs41$ per pixel. The CCD format was
$1024\times1024$ square pixels, giving a field of view of
7.0$\times$7.0 arcmin$^2$.

The raw data frames were first bias subtracted and trimmed using the
standard procedures within IRAF. Pixel-to-pixel sensitivity variations
were then removed by dividing each frame by a normalized high
signal-to-noise mean flatfield

The calibration of the raw photometry was accomplished in the same way
as in Rosenberg et~al (1997). Exposures in each filter of 15
standard stars from Landolt (1992) were taken, allowing a total of
$\sim$50 individual measures per night. The nights were photometric,
and a good calibration of the data was obtained. The total zero--point
errors are of the order 0.01~mag in both filters.

Figure~\ref{m2_cmd} shows the calibrated CMD for the cluster M2, and
illustrates the photometric quality. The RGB and HB are clearly
defined and well sampled. This allows an easy selection of the target
stars for the spectral analysis. They have been chosen on the basis of
the following criteria.
\begin{itemize}
\item luminosity in the same range as the stars to be used for Palomar~1;
\item proximity to the red giant branch;
\item sufficient distance from the cluster center to avoid contamination of the
spectra;
\item possibility to put two stars of similar luminosity into the slit
simultaneously. This optimized the available telescope time.
\end{itemize}

These four criteria were met by the stars marked in Figures~\ref{m2_cmd},
\ref{m2_im}, \ref{m15_im} and \ref{m71_im}. Stars I, II, III and IV in
Figure~\ref{m2_cmd} are all on the RGB, have the same luminosity within
0.7~mag to 1.5~mag below the HB. Looking at Figure~\ref{m2_im} it is
also clear that they are located at $> 4'$ outside the cluster center,
and that their separation allows simultaneous observations in pairs.
A similar selection has been applied to the stars of M15
(Figure~\ref{m15_im}) and M71 (Figure~\ref{m71_im}).

\newpage

\figcaption[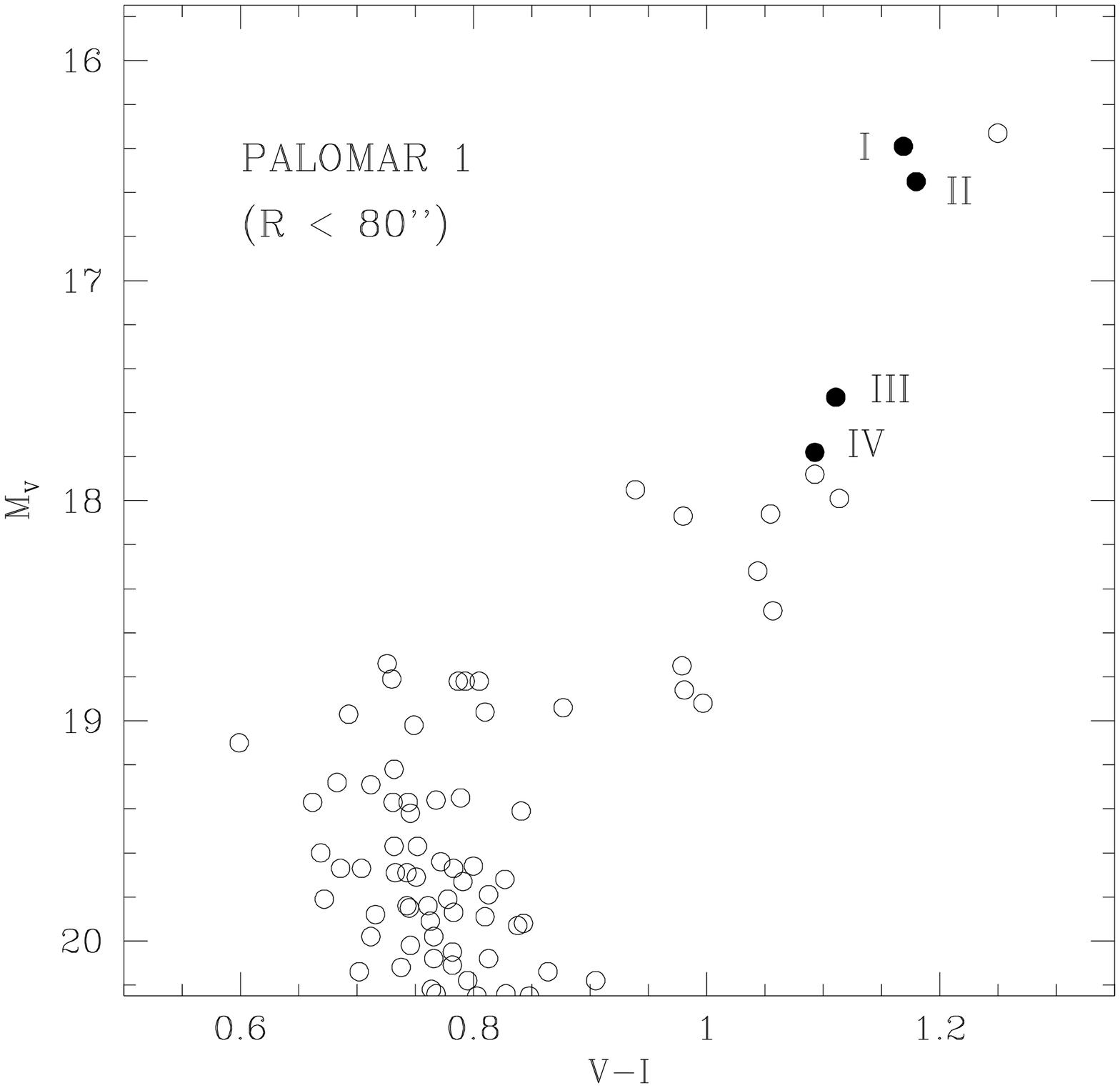]{$V-I$ color-magnitude diagram of Palomar~1. 
Only stars within the first 80 arcsec are represented. The upper part
of the mean sequence and the giant branch are clearly defined. Stars
for which spectra have been obtained are identified by roman numbers.
\label{sel_cmd}}

\figcaption[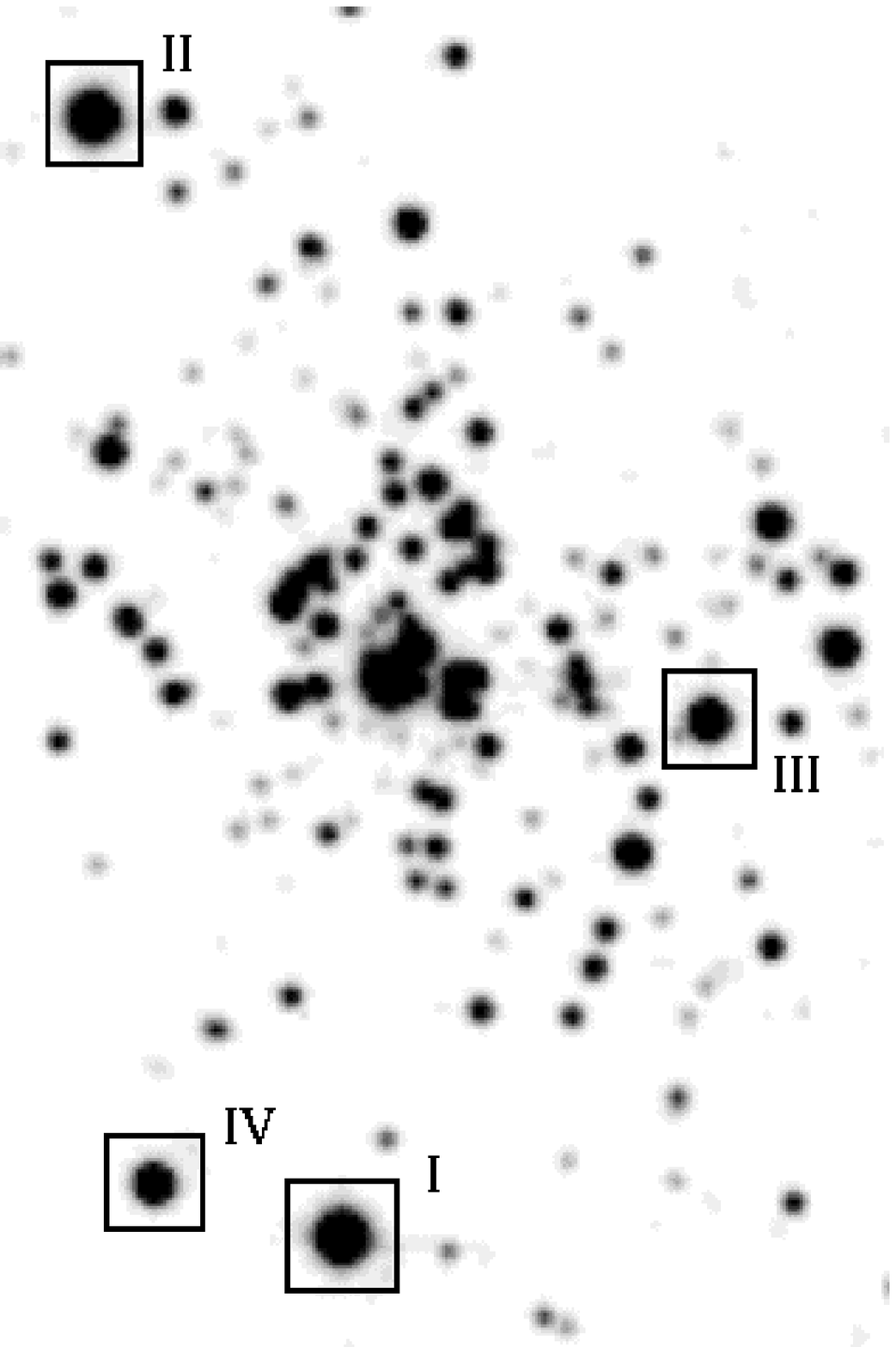]{900s $I$ image of the central part 
of the cluster Pal~1 ($1.4 \times 1.5$ arcmin$^2$, north-up,
west-left). The stars for which spectra have been obtained are marked.
\label{sel_im}}

\figcaption[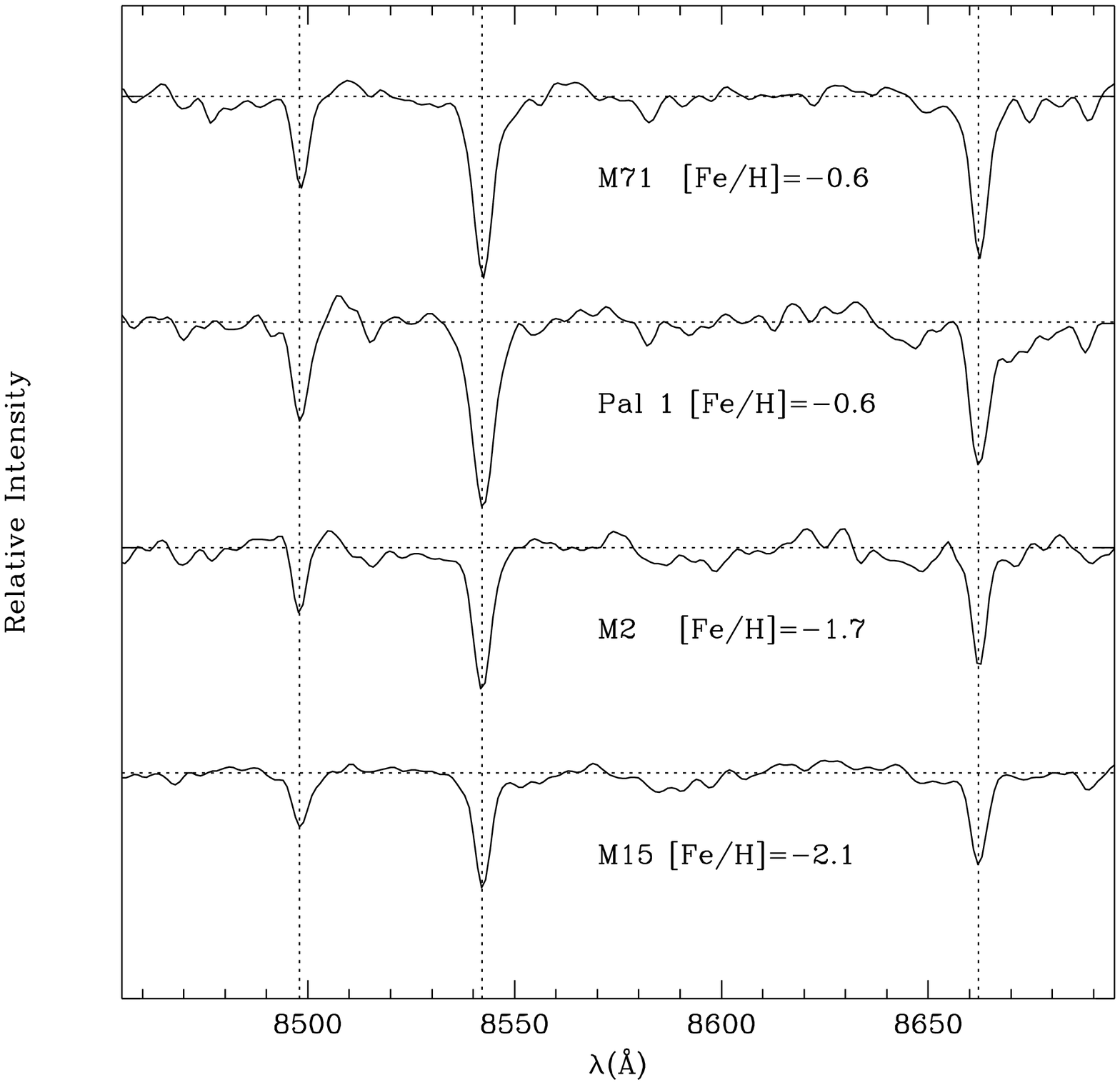]{Sections of the spectra covering 
the Ca~II triplet region for a single star of each observed cluster.
\label{spec}}

\figcaption[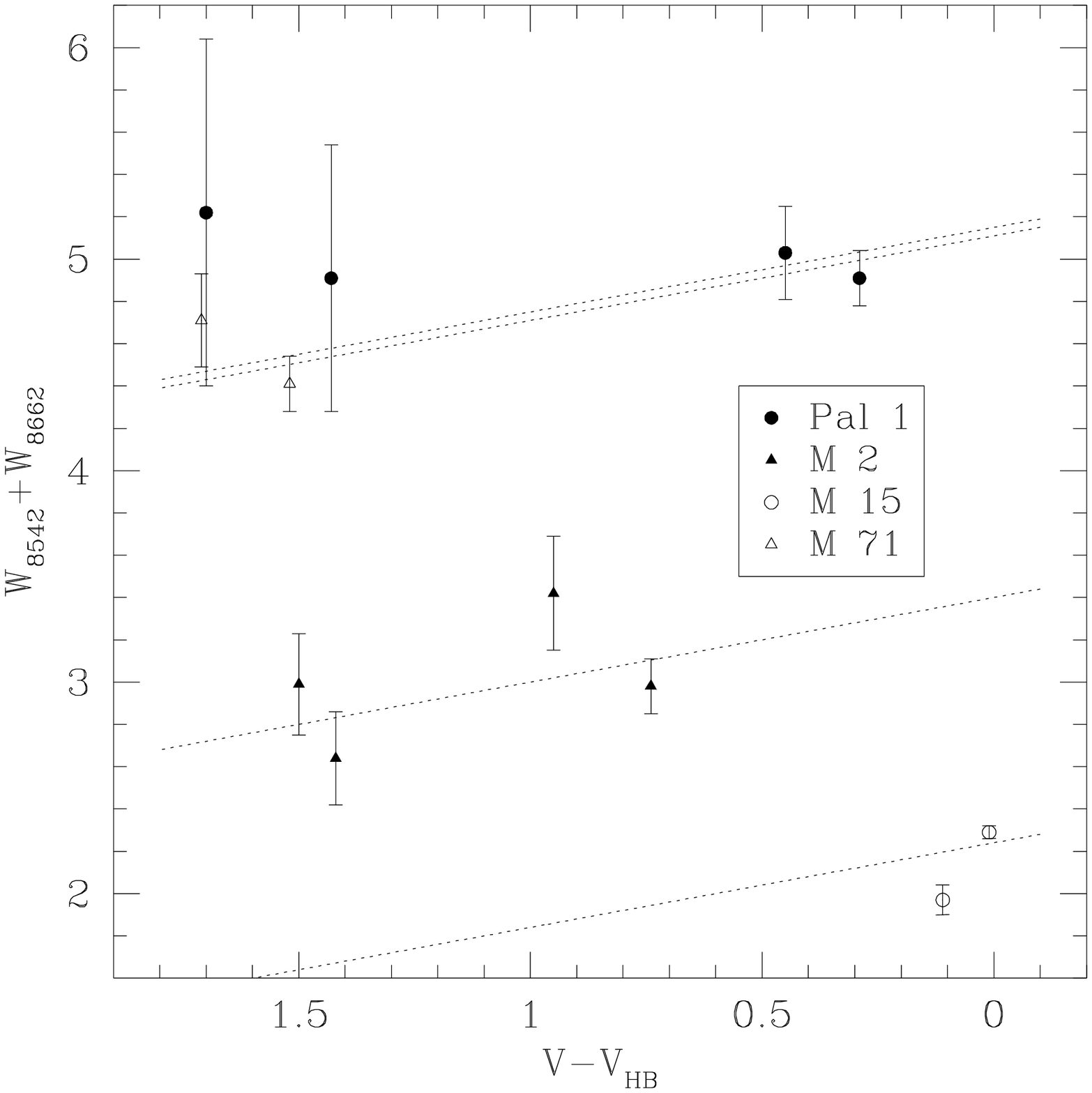]
{The summed equivalent widths and their associated errors are plotted 
vs. $V-V_{HB}$ for the cluster member stars (including stars III and IV of 
M2). Also, dashed lines reproduce the adopted best fit, as described in the 
text.\label{EW}}

\figcaption[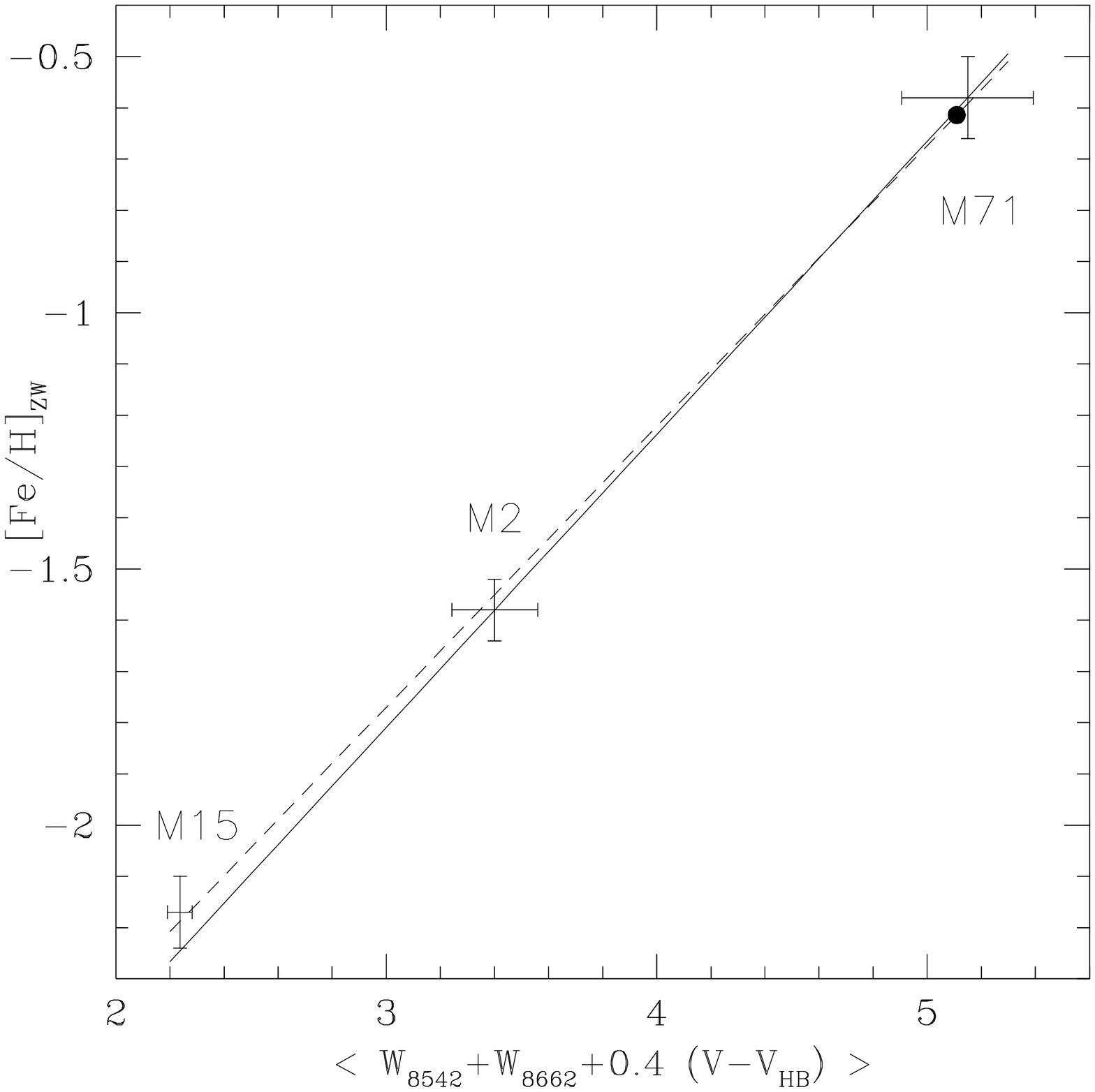]{The abundance calibration for the Ca~II 
line strengths: the reduced equivalent width $W'$ is plotted against
[Fe/H] on the Zinn \& West (1984) scale for the 3 calibration
clusters. The dashed line was fit to the three calibration points by
least squares, while the solid line was fit to M2 and M71. Both lines
represent the adopted calibration relations.\label{met}}

\figcaption[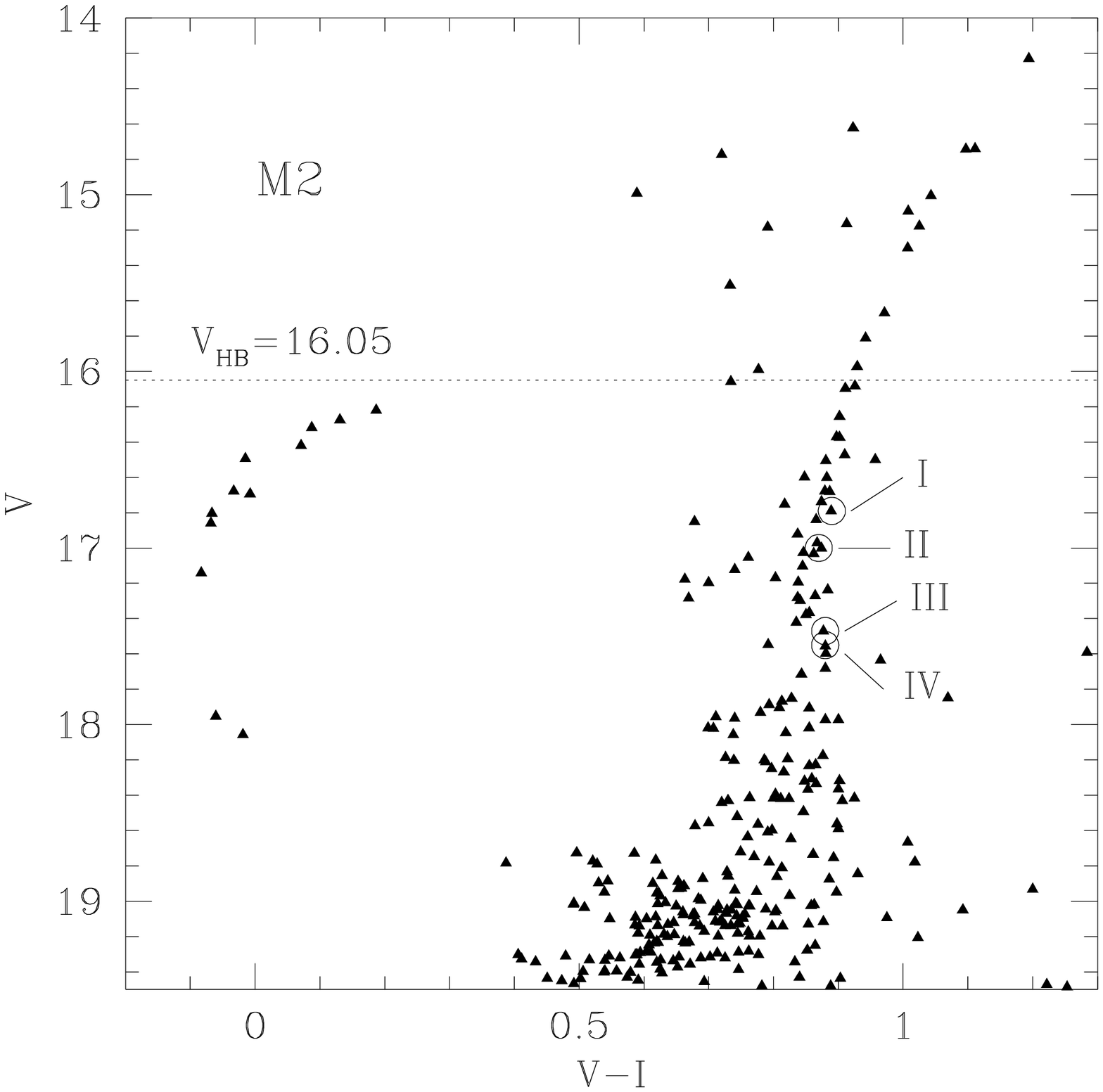]{$V vs. (V-I)$ color-magnitude diagram of M2 
obtained at the IAC-80 telescope. The stars used for spectroscopic
observations are numbered with the same notation used in
Table~2, and illustrate the selection criterium. Stars
close to the RGB, and fainter than the HB were used, matching the same
luminosity range covered by Palomar~1 stars.\label{m2_cmd}}

\figcaption[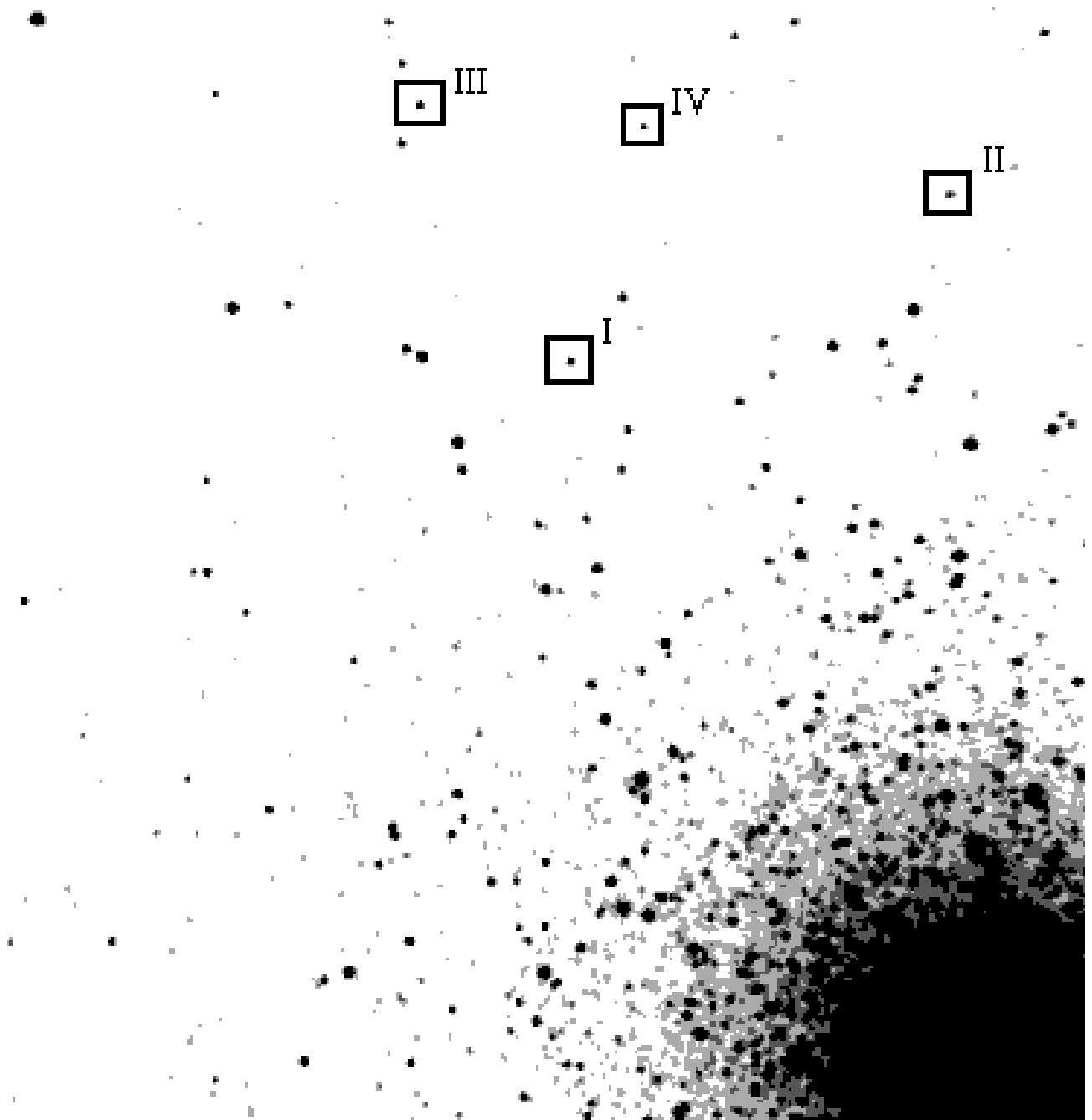]{900s $I$ image of $5 \times 5$ square 
arcmin of M2, showing the observed stars (North is up, West is left). 
Again, the target stars are marked and numbered.\label{m2_im}}

\figcaption[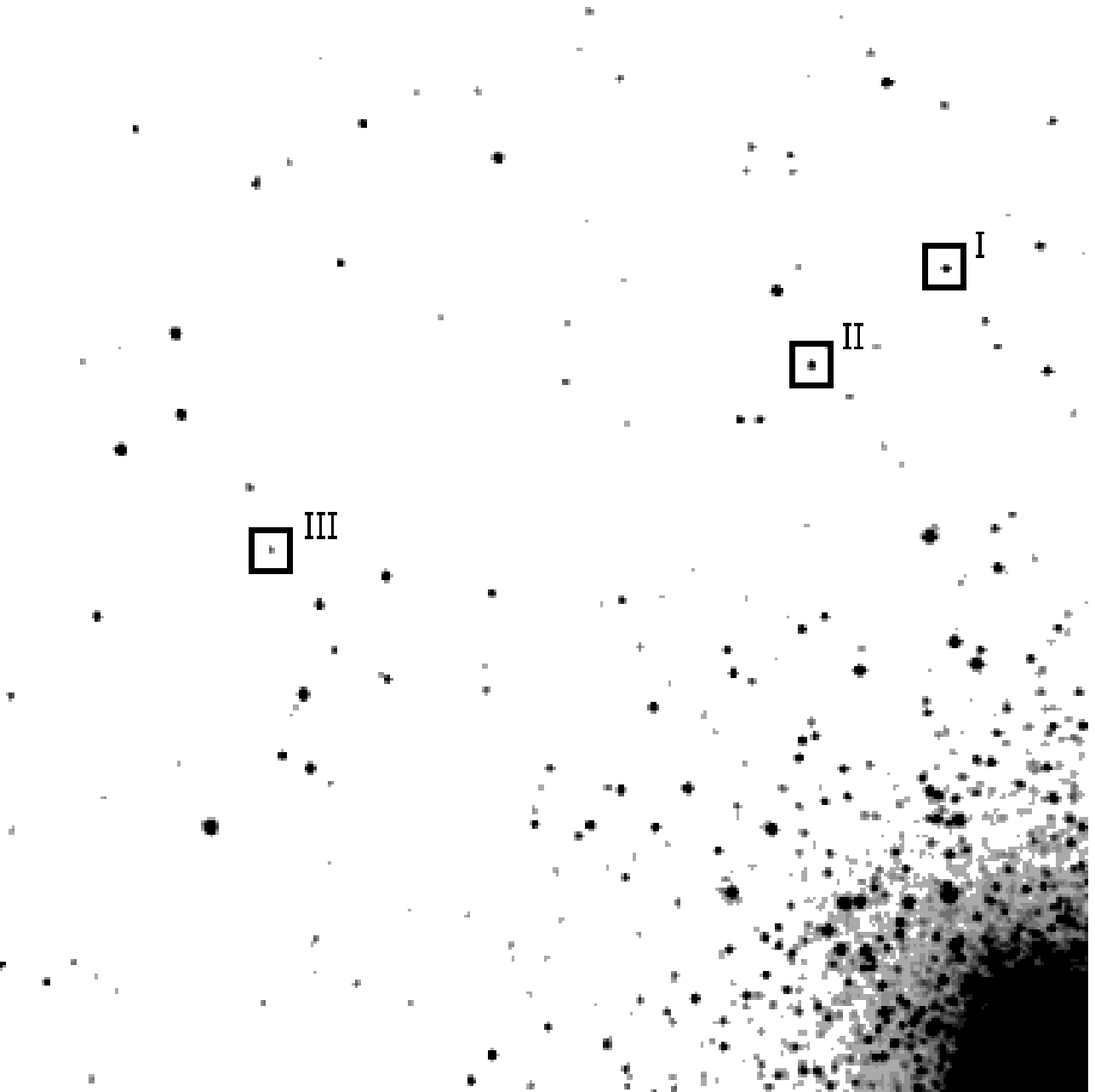]{As in Fig~7 for M15.\label{m15_im}}

\figcaption[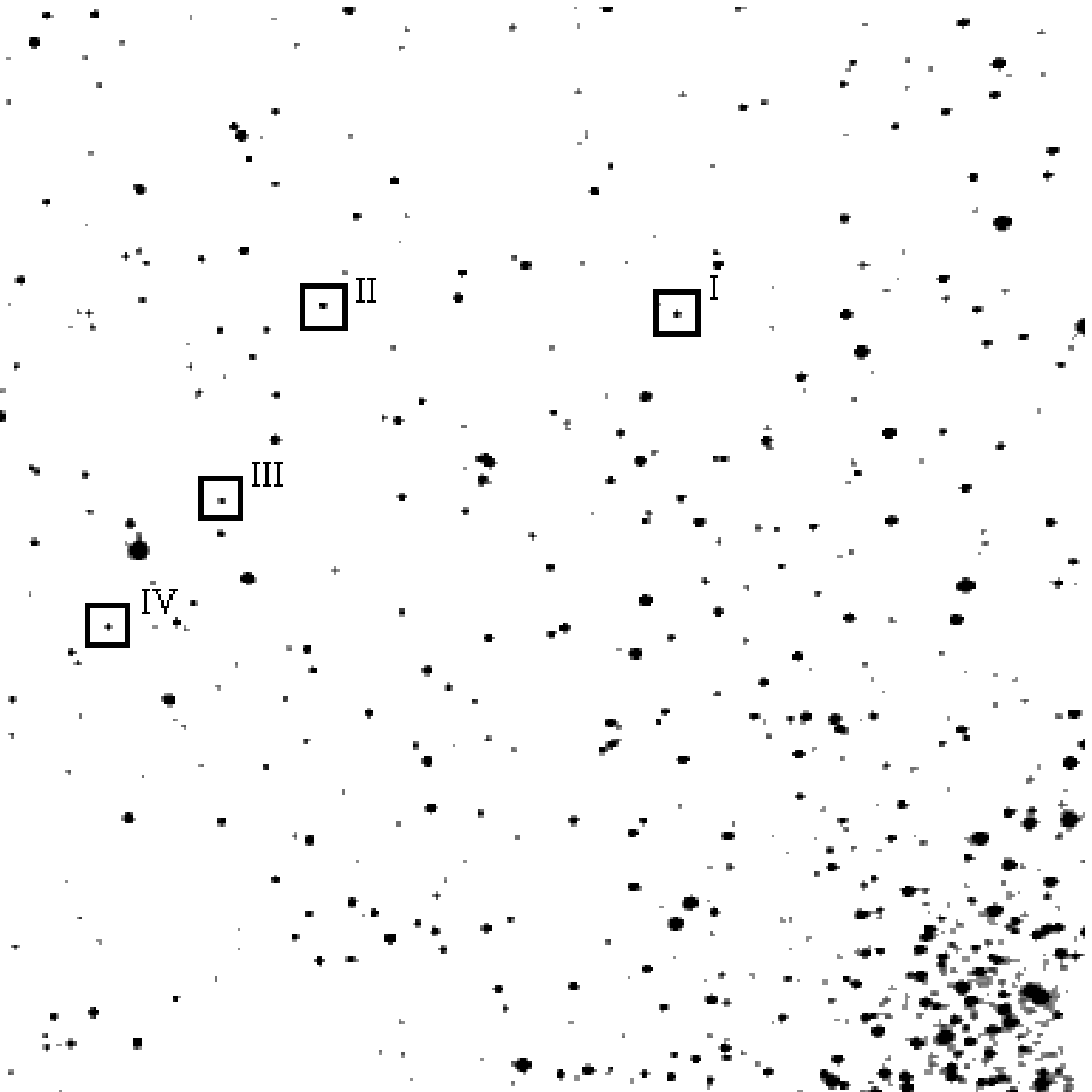]{As in Fig~7 for M71.\label{m71_im}}
 
\dummytable\label{tab:obs}
\dummytable\label{tab:dat}
\dummytable\label{tab:fehpal1}

\end{document}